\documentclass[doublecol]{epl2}
\usepackage{graphicx}
\usepackage{dcolumn}
\usepackage{bm}
\usepackage{epsf}
\usepackage{subfigure}
\usepackage{epstopdf}
\usepackage{amsmath}
\usepackage{amssymb}

\newcommand{\mean}[1]{\langle #1 \rangle}

\newcommand{\e}{{\rm e}}
\newcommand{\nn}{\nonumber\\}

\begin{document}

\title{Stochastic thermodynamics for ''Maxwell demon'' feedbacks}
\shorttitle{Stochastic thermodynamics for ''Maxwell demon'' feedbacks}


\author{Massimiliano Esposito \inst{1} \and Gernot Schaller \inst{2}}
\institute{
 \inst{1} Complex Systems and Statistical Mechanics, University of Luxembourg, L-1511 Luxembourg, Luxembourg.\\
 \inst{2} Institut f\"ur Theoretische Physik, Technische Universit\"at Berlin, Hardenbergstrasse 36, D-10623 Berlin, Germany
}


\abstract{
We propose a way to incorporate the effect of a specific class of feedback processes into stochastic thermodynamics. These ''Maxwell demon'' feedbacks do not affect the system energetics but only the energy barriers between the system states (in a way which depends on the system states). They are thus of a purely informational nature. We show that the resulting formalism can be applied to study the thermodynamic effect of a feedback process acting on electron transfers through a junction.    
}

\maketitle
\section{Introduction}

Our understanding of nonequilibrium statistical mechanics has significantly improved over the last fifteen years, in large part due to our ability to accurately manipulate small fluctuating systems operating far from equilibrium \cite{JarzynskiRev11, BustamanteRitortPD}. At the theoretical level, the discovery of fluctuation theorems (see the reviews \cite{CampisiHanggiTalknerRev, EspositoReview, Harris07} and references therein) and the fundamental new accomplishments in stochastic thermodynamics \cite{VanDenBroeckST86, SeifertST08, Sekimoto10, Esposito12, Seifert12Rev} have played a major role in this respect. In view of these developments, it is therefore not so surprising that we are witnessing a regained interest in the study of the intricate connections between information and thermodynamics \cite{QianKim07, SagawaUedaPRL08, SagawaUedaPRL09, SagawaUedaPRL10, EspositoJSM10, Segal10, Ponmurugan10, Horowitz10, HorowitzParrondo11, EspoVdB_EPL_11, SeifertAbreu11, SeifertAbreuPRL12, SeifertBauerAbreu12, Jayannavar12} recognized long ago by pioneers such as Szilard, Landauer, Bennett, and many others (most of these early works can be found in Ref. \cite{Leff}). What seemed rather abstract and unrealistic considerations have nowadays become experimentally relevant questions \cite{Leigh, ToyabeSagawaUedaNP10, VdBNatPhys, LutzNat12}. This is particularly true when describing systems undergoing feedback processes. In this paper, we propose to incorporate the effect of a specific class of feedback process (which we call ''Maxwell demon'' feedbacks) in the formalism of stochastic thermodynamics and analyze its consequence on the study of thermodynamic efficiencies.  

In stochastic thermodynamics, any system described by a Markovian stochastic dynamics can be shown to satisfy a ''mathematical'' second law of thermodynamics. This means that the Shannon entropy of the system can be separated into two contributions, an always positive contribution which only vanishes when detailed balance is satisfied (i.e., all probability currents between system states vanish), and a remaining entropy flow contribution. Establishing the first law requires the key assumption of {\it local detailed balance}: in its simplest form the logarithm of the ratio between a forward and backward transition rate due to a given reservoir ${\nu}$ is given by the energy difference between the states involved in the transition in units of $k_b T_{\nu}$. This translates the fact that the mechanisms generating the transitions between system states are external reservoirs at equilibrium. Under these circumstances, the entropy flow can be directly connected to the energy flows in the system and the ''mathematical'' second law becomes the physical second law of thermodynamics. In this paper, we show that the local detailed balance assumption can be modified to account for the effect of a specific class of feedback processes which do not affect the energetics of the system, but only modify the energy barriers between system states. Ideally, such feedbacks do not require any work to function since the energy for crossing the barriers is provided by the reservoirs. They only use the ''information'' or ''knowledge'' of the state of the system to adjust the energy barriers accordingly. Thanks to the modified notion of local detailed balance, systems subjected to such feedbacks can still be analyzed within the powerful framework of stochastic thermodynamic and the effect of these feedbacks on the thermodynamic properties of the system can be systematically investigated. Such a theory is particularly useful to differentiate between general and system specific features. We apply our formalism to the study of an electronic Maxwell demon model proposed in Ref. \cite{BrandesSchallerPRB11}. In the spirit of previous studies on the efficiency of small devices operating far from equilibrium \cite{VandenBroeckPRL05, SeifertSchmiedlEPL08, Izumida08, Tu08, EspoLindVdB_EPL09_Dot, EspositoPRL09, EspositoJSM10, SeifertPRL11, Izumida11, ImparatoPelitiEPL12}, we study the efficiency with which the feedback can generate heat or matter fluxes in directions forbidden by the second law in absence of feedbacks.

\section{Stochastic dynamics with ''Maxwell demon'' feedback}

We consider a system in contact with various reservoirs $\nu$ at fixed temperature $T_{\nu}$ and chemical potential $\mu_{\nu}$. The discrete system states are denoted by $m$ and have an energy $\epsilon_m$ and number of particles $N_m$. Transitions betweens systems states are triggered by the reservoirs. The probability for the system to be on the state $m$ evolves according to the Markovian master equation
\begin{eqnarray}\label{MEq}
\dot{p}_m = \sum_{m'} W_{mm'} p_{m'}.
\end{eqnarray}
The rates satisfy $\sum_m W_{mm'}=0$ (due to probability conservation) and contain contributions from different reservoirs $\nu$: $W_{mm'}=\sum_{\nu} W_{mm'}^{(\nu)}$. 
We assume that the system is subjected to a class of feedbacks that do not affect the energetics of the system but only its kinetic properties. 
These ''Maxwell demon feedbacks'' are assumed to modify the local detailed balance property of the rates in the following way 
\begin{eqnarray}\label{ModifiedDB}
\ln \frac{W_{mm'}^{(\nu)}}{W_{m'm}^{(\nu)}}= -\frac{(\epsilon_m - \epsilon_{m'}) - \mu_{\nu} (N_m - N_{m'})}{k_b T_{\nu}} + f_{mm'}^{(\nu)}.
\end{eqnarray}
The feedback parameters $f_{mm'}^{(\nu)}$ cannot depend on $\epsilon_m$ and $N_m$ and are such that $f_{mm}^{(\nu)}=0$ and $f_{mm'}^{(\nu)}=-f_{m'm}^{(\nu)}$.
In absence of feedback, $f_{mm'}^{(\nu)}=0$, we recover the standard local detailed balance property which is generic for system 
interacting with fast equilibrium reservoirs and is known to lead to a consistent thermodynamic description of the system \cite{EspoVdB10_Da}.
The form (\ref{ModifiedDB}) implicitly assumes that the feedback acts by controlling some physical parameters present in the rates 
on timescales much faster than the system and much slower than the reservoirs. It imposes a weak constraint on the explicit form 
of the rates which can assume very different form depending on the system under consideration.
To fix our ideas, let us consider as a first example Arrhenius rates
\begin{eqnarray}\label{Arrhenius}
W_{mm'}^{(\nu)}= A \exp{\{-\frac{B_{mm'}^{(\nu)}-E_{m'}}{k_b T^{(\nu)}}\}},
\end{eqnarray}
where $B_{mm'}^{(\nu)}$ is the energy barrier between state $m'$ and $m$ when the transition is caused by the reservoir $\nu$. We see that the feedback parameter turns out to be the differences between the energy barriers from state $m$ to $m'$ and from $m'$ to $m$  
\begin{eqnarray}\label{FforArrhenius}
f_{mm'}^{(\nu)}=\frac{B_{m'm}^{(\nu)}-B_{mm'}^{(\nu)}}{k_b T^{(\nu)}}.
\end{eqnarray}
\begin{figure}[ht]
\centering
\rotatebox{0}{\scalebox{0.45}{\includegraphics{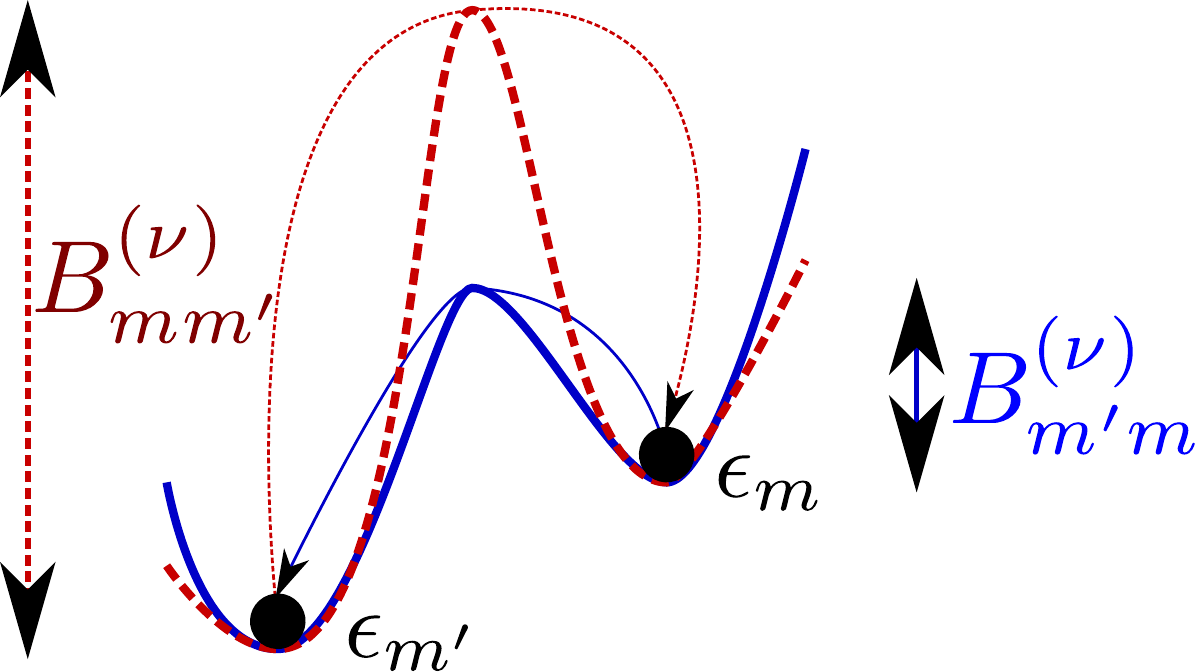}}}
\caption{(Color Online) Illustration of how a feedback, by modifying the potential landscape separating two well defined potential wells, could lead to a discrete description in term of Arrhenius rates (\ref{Arrhenius}). The sketched full (dashed) potential landscape corresponds to transitions described by $W_{mm'}$ ($W_{m'm}$).}
\label{Plot1}
\end{figure}
In absence of feedback, $f_{mm'}^{(\nu)}=0$ because the energy barrier have to be the same in both directions. The feedback acts exclusively on the energy barriers and leaves therefore the system energies unaltered (see Fig. \ref{Plot1}). We call these feedbacks ''Maxwell demon feedbacks'' because they can be seen picturesquely as resulting from a Maxwell demon which is able to instantaneously tune the values of the energy barriers whenever at least one given transition occurs. As a second example we consider ''Fermi golden rule'' rates resulting from a weak interaction between the system and fermionic or bosonic reservoirs at equilibrium. Assuming (without loss of generality) that for fermions $N_{m'}=N_m-1$ and for bosons $\epsilon_{m'}<\epsilon_m$, one obtains
\begin{eqnarray}\label{Fermigoldenrule}
W_{mm'}^{(\nu)}&=& \Gamma_{mm'}^{(\nu)} \; n\left(\frac{\epsilon_m - \epsilon_{m'} - \mu_{\nu}}{k_b T^{(\nu)}}\right)\,,\nn
W_{m'm}^{(\nu)}&=& \Gamma_{m'm}^{(\nu)} \; \left[1 \mp n\left(\frac{\epsilon_m - \epsilon_{m'} - \mu_{\nu}}{k_b T^{(\nu)}}\right)\right]\,,
\end{eqnarray}
where depending on the particle species, $n$ denotes the Fermi or Bose distribution $n(x)=(\e^{x} \pm 1)^{-1}$ (in the latter case the chemical potentials vanish), respectively, and the $\Gamma_{m'm}^{(\nu)}$'s are related to tunneling amplitudes between states. In absence of feedback, these are symmetric $\Gamma_{mm'}^{(\nu)}=\Gamma_{m'm}^{(\nu)}$. The feedback process consists again in modifying the tunneling amplitudes depending on the state of the system. Using~(\ref{ModifiedDB}), we find that the feedback parameters are expressed in terms of the tunneling rates as
\begin{eqnarray}\label{FforQuantum}
f_{mm'}^{(\nu)}= \ln \frac{\Gamma_{mm'}^{(\nu)}}{\Gamma_{m'm}^{(\nu)}} .
\end{eqnarray}

\section{Stochastic thermodynamics with feedback}

We now consider the stochastic thermodynamic description of a system subjected to the above described ''Maxwell demon feedbacks''.
We are going to show that, in contrast to a standard thermodynamic forces, such feedbacks do not affect the first law of thermodynamics but do enter the second law. 
The energy and number of particles in the system are given by 
\begin{eqnarray}\label{FirstLaw}
E=\sum_m p_m \epsilon_m \ \ \;, \ \ N=\sum_m p_m N_m .
\end{eqnarray}
Since the total energy and number of particles are conserved and since ''Maxwell demon feedbacks'' do not affect the system energies and number of particles, 
the energy and matter balance reads  
\begin{eqnarray}\label{EnergyBalance}
\dot{E} = \dot{\lambda} \partial_{\lambda} E + \sum_{\nu} I_{E}^{(\nu)} \ \ \;, \ \
\dot{N} = \dot{\lambda} \partial_{\lambda} N + \sum_{\nu} I_{M}^{(\nu)} .
\end{eqnarray}
The first contribution on the right hand side accounts for changes induced by an external work source whose effect on the system energies, 
$\epsilon_m(\lambda)$, and number of particles, $N_m(\lambda)$ (this latter case seems however unlikely), is parametrized by $\lambda$.
The second contribution accounts for the energy and matter currents entering the system from reservoir $\nu$ 
\begin{eqnarray}\label{HMcurrent}
I_{E}^{(\nu)} &=& \sum_{m,m'} W_{mm'}^{(\nu)} p_{m'} \big( \epsilon_m - \epsilon_{m'} \big)\,,\nn
I_{M}^{(\nu)} &=& \sum_{m,m'} W_{mm'}^{(\nu)} p_{m'} \big( N_m - N_{m'} \big) \label{Currents}.
\end{eqnarray}
The energy balance can be rewritten as the first law of thermodynamics 
\begin{eqnarray}\label{FirstLawBis}
\dot{E} = \dot{{\cal W}} + \sum_{\nu} \dot{{\cal Q}}^{(\nu)},
\end{eqnarray}
where the work flow contains a mechanical and chemical component
\begin{eqnarray}\label{Work}
\dot{{\cal W}} = \dot{\lambda} \partial_{\lambda} E + \sum_{\nu} \mu_{\nu} I_{M}^{(\nu)}
\end{eqnarray}
and where the heat flow with reservoir $\nu$ is given by
\begin{eqnarray}\label{Heat}
\dot{{\cal Q}}^{(\nu)} = I_{E}^{(\nu)} - \mu_{\nu} I_{M}^{(\nu)}.
\end{eqnarray}
The crucial result up to this point is that the first law remains unaffected by the feedback.
We now turn to the system entropy which is given by the Shannon entropy
\begin{eqnarray}\label{ShannonEnt}
S = - k_b \sum_{m} p_{m} \ln p_{m} .
\end{eqnarray}
The entropy balance reads 
\begin{eqnarray}\label{ModEntBal}
\dot{S} = \dot{S}_{\bf e} + \dot{S}_{\bf i} ,
\end{eqnarray}
where the entropy production is given by
\begin{eqnarray}\label{ModEP}
\dot{S}_{\bf i} \equiv k_b \sum_\nu \sum_{m,m'} W_{mm'}^{(\nu)} p_{m'} \ln \frac{ W_{mm'}^{(\nu)} p_{m'} }{ W_{m'm}^{(\nu)} p_{m} } \geq 0
\end{eqnarray}
and the entropy flow by
\begin{eqnarray}\label{ModEF}
\dot{S}_{\bf e} \equiv - k_b \sum_\nu \sum_{m,m'} W_{mm'}^{(\nu)} p_{m'} \ln \frac{W_{mm'}^{(\nu)}}{W_{m'm}^{(\nu)}} .
\end{eqnarray}
Using the modified local detailed balance property~(\ref{ModifiedDB}), this latter can be expressed as
\begin{eqnarray}\label{ModEFbis}
\dot{S}_{\bf e} =  \sum_{\nu} \frac{\dot{{\cal Q}}^{(\nu)}}{T_{\nu}} - I_{F} .
\end{eqnarray}
The first term on the right hand side is the standard form of the entropy flow in absence of feedback. The second term is the information current due to the feedback and reads 
\begin{eqnarray}\label{InfoCurr}
I_{F}=\sum_{\nu} I_{F}^{(\nu)} \ \ , \ \
I_{F}^{(\nu)} = k_b \sum_{m,m'} W_{mm'}^{(\nu)} p_{m'} f_{mm'}^{(\nu)} \label{InfoCurrent}.
\end{eqnarray}
Obviously, while ''Maxwell demon feedbacks'' do not affect the energy and matter balance they do affect the entropy balance.
Using~(\ref{ModEFbis}), we can rewrite~(\ref{ModEntBal}) as
\begin{eqnarray}\label{ebNEW}
\dot{S}_{\bf i} = \dot{S} -\sum_{\nu} \frac{\dot{{\cal Q}}^{(\nu)}}{T_{\nu}} + I_{F} \geq 0.
\end{eqnarray}
\begin{center}
\begin{table*}
\centering
\rotatebox{0}{\scalebox{0.40}{\includegraphics{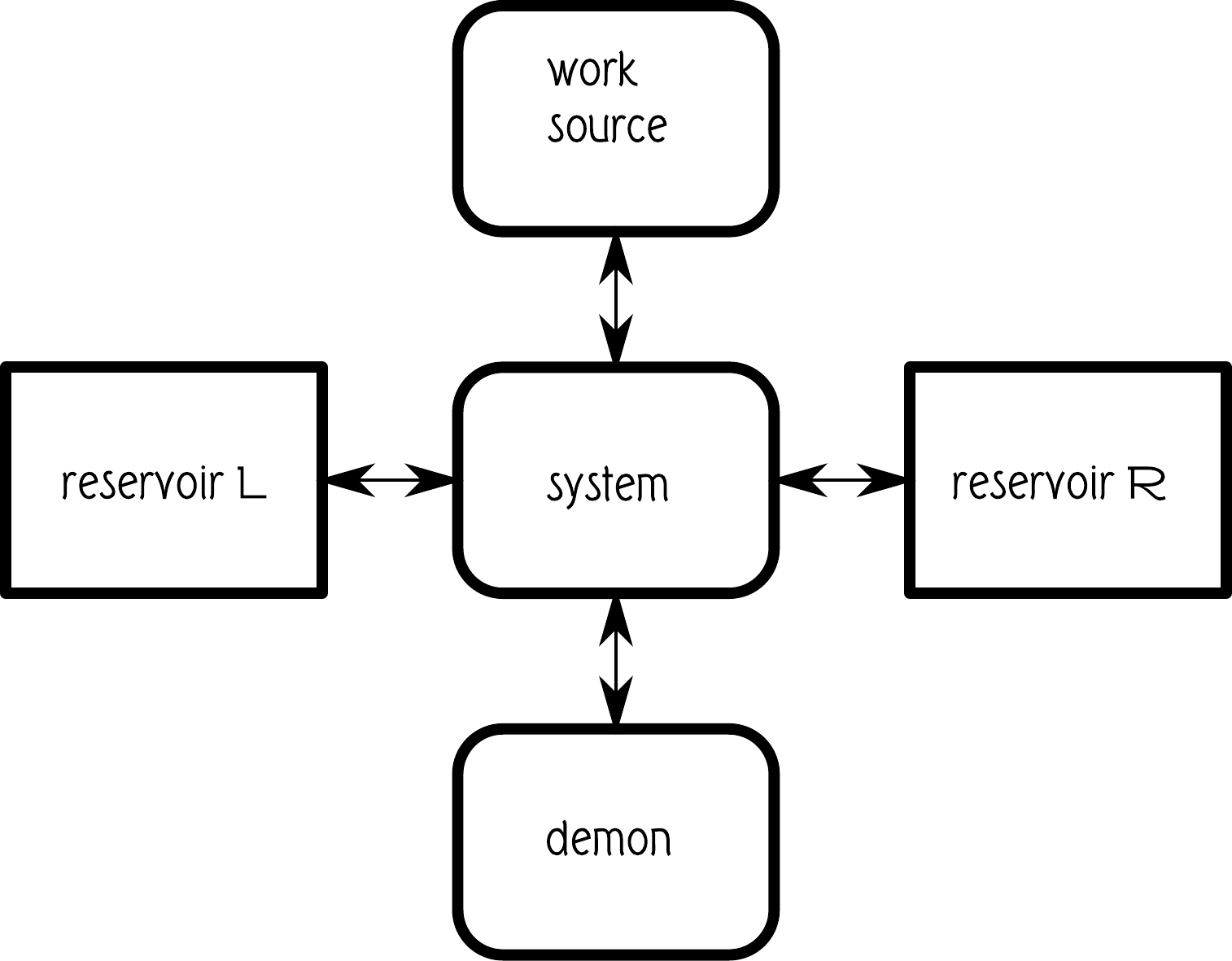}}}\\
\vspace{0.2cm}
\begin{tabular}{|c|c|c|c|c|}
\hline
          & System         & Work source                                            & Reservoir $\nu$                                         & Demon               \\
\hline
Energy    & $\dot{E}$      & $\dot{E}_{\cal W}=-\dot{\lambda} \partial_{\lambda}E$  & $\dot{E}_{\nu}=-I_{E}^{(\nu)}$                          & $\dot{E}_{D}=0$      \\ 
Matter    & $\dot{N}$      & $\dot{N}_{\cal W}=-\dot{\lambda} \partial_{\lambda}N$  & $\dot{N}_{\nu}=-I_{M}^{(\nu)}$                          & $\dot{N}_{D}=0$      \\ 
Entropy   & $\dot{S}$      & $\dot{S}_{\cal W}=0$                                   & $\dot{S}_{\nu}=-\dot{{\cal Q}}^{(\nu)}/T_{\nu}$         & $\dot{S}_{D}=-I_{F}$ \\
\hline
\multicolumn{5}{|l|}{Total energy conservation: $\dot{E}+\dot{E}_{\cal W}+\sum_{\nu} \dot{E}_{\nu}=0$ $\ \ \rightarrow \ \ $ Eq. (\ref{EnergyBalance}) } \\
\multicolumn{5}{|l|}{Total matter conservation: $\dot{N}+\dot{N}_{\cal W}+\sum_{\nu} \dot{N}_{\nu}=0$ $\ \ \rightarrow \ \ $ Eq. (\ref{EnergyBalance}) } \\
\multicolumn{5}{|l|}{Total entropy production:  $\dot{S}_{\bf i} = \dot{S}+\sum_{\nu} \dot{S}_{\nu}+\dot{S}_{D} \geq 0$ $\ \ \rightarrow \ \ $ Eq. (\ref{ebNEW}) } \\
\hline
\end{tabular}
\caption{Illustration of each element constituting the total system. The table lists their respective energy, matter and entropy change.}
\label{my_table}
\end{table*}
\end{center}
This is a central result of this paper. In absence of feedback, $I_{F}=0$, it is well known that entropy production can be interpreted as a ''total entropy'' because it can be seen as the sum of the entropy change in the system, $\dot{S}$, and the entropy changes in the reservoirs. Indeed, the entropy change in an ideal (i.e., reversible) reservoir $\nu$ is given by the heat flowing into it divided by its temperature, i.e., $\dot{S}_{\nu}=-\dot{{\cal Q}}^{(\nu)}/T_{\nu}$. As a result, the positivity of $\dot{S}_{\bf i}$ implies that $\dot{S} \geq \sum_{\nu} \dot{{\cal Q}}^{(\nu)}/T_{\nu}$ which is the traditional second law of thermodynamics. In presence of feedback, depending on the sign of $I_{F}$ this result need not hold anymore. 
The entropy production $\dot{S}_{\bf i}$ can still be interpreted as the ''total entropy'', but in addition to the change in the entropy of the system and the reservoirs, it also needs to contain the entropy change provided by the feedback mechanism, $I_{F}$. The notion of equilibrium is always defined by $\dot{S}_{\bf i} = 0$, since it still corresponds to the situation where detailed balance is satisfied and where as a result all currents vanish $I_{E}=I_{M}=I_{F}=0$ \cite{KampenB97}. 
The results obtained so far are summarized in Table \ref{my_table} in order to facilitate their interpretation. Each element constituting the total system is subjected to a given change in energy, matter and entropy. We clearly see that while ''Maxwell demon feedbacks" do not modify the energy and matter balance, they do affect the entropy balance. This summary also reveals an interesting duality between the work source and the ''Maxwell demon". While the former is an idealized source of energy without associated entropy generation, the latter is an idealized source of entropy without associated energy changes. 

Without going into details which have been often reported elsewhere (see e.g. \cite{EspositoVdBPRL10} or \cite{Seifert05}), it is clear that the dynamics we considered implies a fluctuation theorem for the entropy production defined at the trajectory level 
\begin{eqnarray}\label{FT}
\mean{\e^{-(\Delta_{\bf i} S)/k_b}} = \mean{\e^{-(\Delta S - \sum_{\nu} {\cal Q}^{(\nu)}/T_{\nu} + F)/k_b}} = 1,
\end{eqnarray}
where $F$ is the integrated information current $I_{F}$ defined at the trajectory level. This integral fluctuation theorem is the analog of the fluctuation theorems derived in \cite{SagawaUedaPRL10} for systems subjected to feedback and in contact with a single reservoir (in this latter case $\Delta S - \sum_{\nu} {\cal Q}^{(\nu)}/T_{\nu}=({\cal W}-\Delta F)/T$). We note that the detailed fluctuation theorem also holds. The backward dynamics is identical to the forward dynamics and is subjected to the same ''Maxwell demon'' feedback as the forward dynamics in contrast to detailed fluctuation theorems obtained for other feedbacks which act by an external (time dependent) control of the rates \cite{Horowitz10}.

From now on we will focus on nonequilibrium steady state situations and consider for simplicity the case of two reservoirs $\nu=L,R$. 
This means that due to energy and matter conservation we have $I_{E,M} \equiv I_{E,M}^{(L)}=-I_{E,M}^{(R)}$. 
Since furthermore in steady state one has $\dot{S}=0$, it follows that $\dot{S}_{\bf i}=-\dot{S}_{\bf e}$ and~(\ref{ebNEW}) becomes
\begin{eqnarray}\label{EP_SS}
\dot{S}_{\bf i} = \left( \frac{1}{T_{R}}-\frac{1}{T_{L}} \right) I_{E} - \left( \frac{\mu_R}{T_{R}}-\frac{\mu_L}{T_{L}} \right) I_{M} + I_{F} \geq 0 .
\end{eqnarray}
In an isothermal system $T \equiv T_{L}=T_{R}$ for example, assuming that $\mu_R \geq \mu_L$, Eq.~(\ref{EP_SS}) becomes
\begin{eqnarray}\label{EP_SSisochem}
\dot{S}_{\bf i} = - \frac{{\cal P}}{T} + I_{F} \geq 0,
\end{eqnarray}
where the extracted power is given by 
\begin{eqnarray}\label{Power}
{\cal P} = -\dot{{\cal W}} = \sum_{\nu} \dot{{\cal Q}}^{(\nu)} = ( \mu_R-\mu_L ) I_{M}.
\end{eqnarray}
In absence of feedback, the matter flux can only flow down the chemical potential gradient (i.e., $I_M \leq 0$). However, in presence of feedback, if $I_{F}$ is sufficiently positive, the matter flux can climb up the chemical gradient (i.e., $I_M \geq 0$) and lead to positive extracted power with an efficiency 
\begin{eqnarray}\label{Efffeedtowork}
\eta = \frac{{\cal P}}{T I_{F}} = 1-\frac{\dot{S}_{\bf i}}{I_{F}}.
\end{eqnarray}
A similar analysis can be done in absence of a chemical potential gradient $\mu \equiv \mu_L=\mu_R$ and assuming that $T_L \geq T_R$. In this case, entropy production~(\ref{EP_SS}) becomes 
\begin{eqnarray}\label{EP_SSisotemp}
\dot{S}_{\bf i} = -\frac{\eta_C}{T_R} \dot{{\cal Q}}^{(R)} + I_{F} \geq 0 ,
\end{eqnarray}
where $\eta_C=1-T_R/T_L$ is the Carnot efficiency. Thanks to the feedback, heat could flow from lower to higher temperature (i.e., $\dot{Q}_R=-\dot{Q}_L>0$) and cool the cold reservoir with an efficiency
\begin{eqnarray}\label{Efffeedtoheat}
\eta = \frac{\dot{{\cal Q}}^{(R)}}{T_R I_{F}} = \frac{1}{\eta_C} \left(1-\frac{\dot{S}_{\bf i}}{I_{F}}\right). 
\end{eqnarray}
As a final example, we mention that the feedback could also be used to improve the efficiency of a thermoelectric generator. To see this, we assume that $T_L \geq T_R$ and $\mu_R \geq \mu_L$ and rewrite~(\ref{EP_SS}) as 
\begin{eqnarray}\label{EP_SSbis}
\dot{S}_{\bf i} = -\frac{{\cal P}}{T_R} + \frac{\eta_C}{T_R} \dot{{\cal Q}}^{(L)} + I_{F} \geq 0 .
\end{eqnarray}
The thermoelectric effect occurs when ${\cal P} > 0$. In absence of feedback, this requires heat from the hot reservoir $\dot{{\cal Q}}^{(L)}>0$. The efficiency of this thermal engine is usually defined by 
\begin{eqnarray}\label{Efficiency}
\eta = \frac{{\cal P}}{\dot{{\cal Q}}^{(L)}} = \eta_C + \frac{T_R(I_F-\dot{S}_{\bf i})}{\dot{{\cal Q}}^{(L)}}\,,
\end{eqnarray}
which, for $I_F=0$, is upper-bounded by $\eta_C$. We easily see that the feedback can be such that this efficiency increases beyond Carnot efficiency. The reason is that the power generation is not only powered by the heat from the hot reservoir but also by the information flow resulting from the feedback. 

\section{Single level quantum dot with feedback}

We now turn to the thermodynamic analysis of a model system, proposed in Ref. \cite{BrandesSchallerPRB11}, which consists of a single level quantum dot in contact with two fermionic reservoirs and subjected to a Maxwell demon feedback. The dot can be empty ($m=0$) or filled ($m=1$) and the rates describing the reservoir-induced transitions between these states are given by 
\begin{eqnarray}\label{ModelRates}
W_{10}^{(\nu)} = \Gamma_{\nu} n_{\nu}(\epsilon) \  \ , \ \
W_{01}^{(\nu)} = \Gamma_{\nu} e^{f_{\nu}} \left[ 1-n_{\nu}(\epsilon) \right] ,
\end{eqnarray}
where the Fermi distribution is given by \mbox{$n_{\nu}(\epsilon)=(e^{x_{\nu}}+1)^{-1}$} and $x_{\nu}=(\epsilon-\mu_{\nu})/(k_b T_{\nu})$.
The local detailed balance condition modified to include the feedback effect therefore reads  
\begin{eqnarray}\label{ModelLDB}
\ln \frac{W_{10}^{(\nu)}}{W_{01}^{(\nu)}} = - x_{\nu} - f_{\nu} .
\end{eqnarray}

\begin{figure}[ht]
\rotatebox{0}{\scalebox{0.55}{\includegraphics{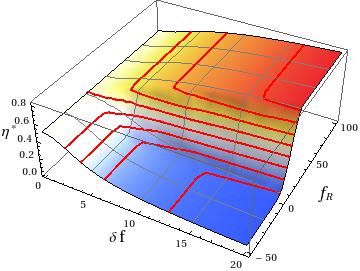}}}
\rotatebox{0}{\scalebox{0.555}{\includegraphics{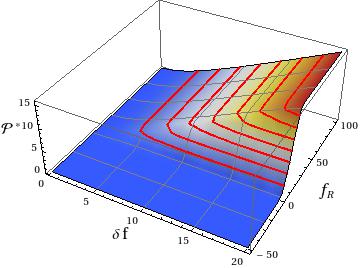}}}
\rotatebox{0}{\scalebox{0.55}{\includegraphics{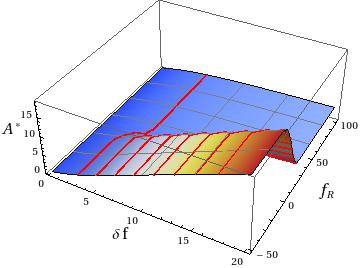}}}
\caption{(Color Online) 
The feedback pumps electrons against the bias (isothermal leads) thus generating power. For maximum power versus $A$ and $x_L$, the efficiency (top), power (middle), and affinity (bottom) of the process is plotted as a function of $\delta f$ and $f_R$. All quantities are dimensionless except power which is expressed in $\Gamma k_b T$ with $\Gamma=\Gamma_{L}=\Gamma_{R}$.}
\label{PlotA&B&C}
\end{figure}

If $p$ denotes the probability to find the dot filled, at steady state we find $p=W_{10}/(W_{10}+W_{01})$.
We introduce the steady state probability current (which for this example is equal to the matter current $I_M$)
\begin{eqnarray}
I &=& W_{10}^{(L)} (1-p) - W_{01}^{(L)} p \nn
&=& \frac{W_{01}^{(L)} W_{10}^{(R)}}{W_{10}+W_{01}} \left( \e^{A} - 1\right) ,
\end{eqnarray}
where we defined the affinity
\begin{eqnarray}\label{ModelAff}
A = \ln \frac{W_{10}^{(L)} W_{01}^{(R)}}{W_{01}^{(L)} W_{10}^{(R)}} = \delta f - \delta x 
\end{eqnarray}
in terms of $\delta f=f_R-f_L$ and $\delta x=x_{L}-x_{R}$. 
The affinity may also be written more explicitly as 
\begin{eqnarray}\label{ModelAffbis}
A = \frac{\epsilon}{k_b} \left( \frac{1}{T_{R}}-\frac{1}{T_{L}} \right) 
+ \frac{1}{k_b} \left( \frac{\mu_L}{T_{L}}-\frac{\mu_R}{T_{R}} \right) + (f_R-f_L) .
\end{eqnarray}

The heat and matter current are proportional in this model $I=I_M=I_E/\epsilon$. This is the so-called tight coupling property. Since $f_{\nu}=f_{01}^{(\nu)}=-f_{10}^{(\nu)}$, the information current~(\ref{InfoCurr}) becomes $I_{F}=k_b \delta f I$ and is thus also proportional to $I$. The three currents (matter, energy and information) are thus tightly coupled. As a result, the entropy production can be written as a single collapsed affinity $A$ times the current $I$:
\begin{eqnarray}\label{EP_SSmodel}
\dot{S}_{\bf i} = k_b A I = k_b (\delta f - \delta x) I\,.
\end{eqnarray}

We will restrict our analysis to the isothermal regime $T \equiv T_R = T_L$, where the feedback is used to generate power by pumping electrons against the bias. In this case the power~(\ref{Power}) can be written as 
\begin{eqnarray}\label{Powermodel}
{\cal P} = k_b T \delta x I = k_b T ( \delta f - A) I 
\end{eqnarray}
and the efficiency for generating this power~(\ref{Efffeedtowork}) becomes
\begin{eqnarray}\label{Efffeedtoworkmodel}
\eta = \frac{\delta x}{\delta f} = 1-\frac{A}{\delta f}.
\end{eqnarray}
At equilibrium, $A = 0$, the efficiency reaches its upper bound $\eta=1$. Since close to equilibrium, the current becomes linear in the affinity, $I=LA$, we observe the well know result that $\eta=1$ corresponds to ${\cal P}=0$. We therefore turn our attention to the efficiency at maximum power with respect to $A$. In the linear (to affinity) regime, the maximum occurs at the affinity $A^*=\delta f/2$ and thus leads to the well known result (for models with tight coupling) that the efficiency at maximum power $\eta^*$ in the linear response regime is half of the ideal equilibrium efficiency, i.e., $\eta^*=1/2$ \cite{VandenBroeckPRL05}. To study the efficiency at maximum power beyond linear response, we need to resort to numerics. Generally, even for equal tunneling rates $\Gamma=\Gamma_L=\Gamma_R$, power in Eq.~(\ref{Powermodel}) will still depend on $x_{L/R}$ and $f_{L/R}$. Using $\delta f = f_R-f_L$, $\delta x=x_L-x_R$, and Eq.~(\ref{ModelAff}), we choose to eliminate these in favor of $\delta f$, $x_L$, and $f_R$. Maximizing the power numerically with respect to both $A$ and $x_L$, the maximum power $P^*$ still depends on $f_R$ and $\delta f$. On figure \ref{PlotA&B&C} we therefore plot, as a function of $\delta f$ and $f_R$, the efficiency, power and affinity corresponding to the power maximum obtained by maximization versus $A$ and $x_L$. We clearly see that close to equilibrium at low affinities (corresponding to small $\delta f$), we recover the universal $1/2$ behavior. We also see that the efficiency at maximum power in the nonlinear regime can become much larger (respectively smaller) than $1/2$ for large (respectively low) values of $f_R$. 


\section{Summary and Outlook}

We considered nonequilibrium systems subjected to ''Maxwell demon'' feedbacks, i.e., feedbacks which do not affect the system energetics but only the energy barriers between system states, and showed that their thermodynamic properties can be studied using the theory of stochastic thermodynamics by extending the traditional local detailed balance as described in Eq. (\ref{ModifiedDB}). We demonstrated that these feedbacks may be used to convert information into work, to cool a cold reservoir, or to increase the standard efficiencies of heat to work conversion above Carnot efficiency (since in this latter case it is actually heat and information in combination that are converted to work). Using a simple model system introduced in \cite{BrandesSchallerPRB11}, we also studied in detail the efficiency at maximum power of information to work conversion. Generalizing the present rate equations-based scheme to quantum master equations (not reducible to rate equations) is an interesting venue for future research.

\acknowledgments

M.~E. is supported by the National Research Fund, Luxembourg in the frame of project FNR/A11/02.
G.~S. gratefully acknowledges support by the DFG (SCHA 1646/2-1).



\end{document}